\documentclass[fleqn,10pt]{wlscirep}
\usepackage[utf8]{inputenc}
\usepackage[T1]{fontenc}

\usepackage{xcolor}
\usepackage{ulem}

\usepackage{floatrow}
\usepackage{graphicx}
\usepackage[label font=bf]{subfig}
\usepackage{caption}
\usepackage{todonotes}

\graphicspath{{Figure/}} 

\title{Computational Advances in Taste Perception: From Ion Channels to Neural Coding}

\author[1]{Vladimir A. Lazovsky}
\author[1,2,*]{Sergey V. Stasenko}
\author[1,2]{Victor B. Kazantsev}
\affil[1]{Moscow Institute of Physics and Technology, Laboratory of neurobiomorphic technologies, Moscow, 117303, Russia}
\affil[2]{Lobachevsky State University of Nizhny Novgorod, Neurotechnology department, 603022, Russia}

\affil[*]{stasenko@neuro.nnov.ru}


\begin{abstract}
Recent advances in computational neuroscience demand models that balance biophysical realism with scalability. We present a hybrid neuron model combining the biophysical fidelity of Hodgkin-Huxley (HH) dynamics for taste receptor cells with the computational efficiency of Izhikevich spiking neurons for large-network simulations. Our framework incorporates biomorphic taste cell models, featuring modality-specific receptor dynamics (T1R/T2R, ENaC, PKD) and Goldman-Hodgkin-Katz (GHK)-driven ion currents to accurately simulate gustatory transduction. Synaptic interactions are modeled via glutamate release kinetics with alpha-function profiles, AMPA receptor trafficking regulated by phosphorylation, and spike-timing-dependent plasticity (STDP) to enforce temporal coding. At the network level, we optimize multiscale learning, leveraging both temporal spike synchrony (van Rossum metrics) and combinatorial population coding (rank-order patterns). This approach bridges single-cell biophysics with ensemble-level computation, enabling efficient simulation of gustatory pathways while retaining biological fidelity.
\end{abstract}
\begin{document}

\flushbottom
\maketitle
%
%
\thispagestyle{empty}


\section*{Introduction}

The gustatory system plays a crucial role in detecting and processing taste stimuli, enabling the perception of five primary taste modalities: sweet, sour, bitter, salty, and umami \cite{chen2022progress}. Taste perception begins when chemical compounds dissolve in saliva and interact with taste buds, which are clusters of specialized sensory cells located primarily on the tongue \cite{lalonde1961number, miller1986variation}. Each taste bud typically contains between 50 and 100 epithelial cells, which can be divided into three main types based on their structural characteristics and functional roles \cite{chaudhari2010cell}. 
These taste cells express specific receptors that recognize different tastants and initiate intracellular signaling pathways. Sweet, bitter, and umami tastes are detected by G protein-coupled receptors (GPCRs), which activate complex signaling cascades involving calcium influx and changes in membrane potential \cite{von2021sweet}. Type I cells, often described as glial-like, provide metabolic and structural support to other taste cells and are implicated in the detection of salty stimuli \cite{ahmad2020g}. Type II cells function as receptor cells, directly responding to sweet, bitter, and umami stimuli through GPCR activation \cite{chaudhari2010cell}. In contrast, Type III cells act as presynaptic cells that primarily respond to sour stimuli and transmit signals to sensory neurons through the release of neurotransmitters such as serotonin and gamma-aminobutyric acid (GABA) \cite{huang2008presynaptic}.

The molecular mechanisms that underlie taste perception differ depending on the specific taste modality. Sweet and umami tastes are detected through heterodimeric receptors belonging to the TAS1R family: the TAS1R2/TAS1R3 combination responds to sweet compounds, while the TAS1R1/TAS1R3 pairing is responsible for sensing umami substances such as L-glutamate \cite{diepeveen2022molecular}. Bitter taste perception is mediated by a diverse group of approximately 30 TAS2R receptors. Some of these, like TAS2R10 and TAS2R14, are broadly tuned to respond to a wide range of bitter compounds, whereas others are more selective and recognize specific bitter molecules \cite{meyerhof2010molecular}. Salty taste detection is believed to involve epithelial sodium channels (ENaCs), although the complete signaling pathway has yet to be fully clarified \cite{vandenbeuch2008amiloride}. Sour taste perception relies on proton-sensitive mechanisms, including the activation of ion channels such as HCN1 and HCN4, as well as the inhibition of certain potassium channels, which contributes to the depolarization of taste cells in response to acidic stimuli \cite{chang2010proton}.

Signal transduction in taste cells, particularly for sweet, bitter, and umami stimuli, follows a well-defined molecular pathway. Activation of the corresponding taste receptors initiates a signaling cascade involving G proteins—most notably gustducin—which subsequently activates phospholipase C$\beta$2. This leads to the production of inositol triphosphate (IP$_3$), which triggers the release of calcium from intracellular stores \cite{banik2021bitter}. The resulting increase in intracellular calcium promotes the release of neurotransmitters, primarily ATP, through the CALHM1 ion channel \cite{taruno2013calhm1}. These chemical signals are then conveyed to the central nervous system via afferent fibers of the facial (VII), glossopharyngeal (IX), and vagus (X) cranial nerves, where they are processed to generate the perception of taste \cite{smith2001making}.

The field of neuromorphic gustatory systems lies at the intersection of biology and engineering, where electronic tongues (e-tongues) emulate the human sense of taste by combining advanced sensor technologies with brain-inspired computational models. Evolving from basic chemical sensor arrays, these systems have developed into highly sophisticated platforms capable of performing real-time, energy-efficient taste analysis. Their applications extend across diverse domains, including food science, healthcare, and environmental monitoring.

At the heart of modern neuromorphic gustatory systems are bioinspired sensor arrays designed to replicate the functionality of biological taste buds. These arrays typically utilize electrochemical sensing techniques—such as potentiometric and voltammetric methods—and are often fabricated using CMOS and MEMS technologies to enable miniaturization and large-scale integration \cite{li2016cmos}. Recent innovations have introduced advanced materials, including conductive polymer nanocomposites and two-dimensional materials, which offer improved sensitivity and selectivity while facilitating seamless integration with neuromorphic computing architectures \cite{han2022bioinspired}. In contrast to conventional electronic tongues that depend on continuous analog signal acquisition, neuromorphic systems employ an event-driven approach in which taste stimuli are encoded as sparse spike trains. This spiking paradigm significantly reduces power consumption and minimizes data redundancy.

The signal processing architecture of neuromorphic gustatory systems is directly inspired by the biological gustatory pathway. Spiking Neural Networks (SNNs) are employed to interpret spike patterns generated by the sensor arrays, enabling in-sensor computing that performs feature extraction and preliminary classification directly at the edge \cite{yang2022neuromorphic}. This biologically inspired, event-driven approach allows for real-time processing with millisecond-level latency, which is essential for time-sensitive applications such as quality control in food production. For higher-level pattern recognition and taste classification, machine learning techniques are integrated into the system. These include unsupervised methods like Principal Component Analysis (PCA) and supervised algorithms such as Artificial Neural Networks (ANNs), which together enhance the system’s ability to accurately interpret complex taste profiles \cite{geladi1986partial,abdilynne2010principal}.

Practical applications highlight the versatility and impact of neuromorphic gustatory systems across multiple domains. In the food industry, these systems facilitate automated quality control by detecting subtle variations in flavor profiles and identifying potential contaminants \cite{lebow2021real}. In healthcare, they support personalized nutrition monitoring, such as tracking sodium intake for individuals with hypertension or monitoring glucose levels for diabetic patients \cite{yang2022neuromorphic}. Environmental monitoring also benefits from this technology, with electronic tongues capable of detecting trace levels of pollutants in water, often at parts-per-billion concentrations \cite{krantz2001electronic}.




\subsection*{Existing Models and Our Approach}

The mathematical modeling of the gustatory system presents a range of challenges and limitations, primarily due to the intricate nature of taste perception and its integration with other sensory modalities. Gustatory sensation arises from chemical signals, whose functions are not yet fully understood. Moreover, taste perception is influenced by factors such as vision, olfaction, thermal sensation, and memory, complicating the development of accurate mathematical representations. The complexity deepens with the integration of multiple sensory inputs—taste, odor, texture, pungency, and temperature—most of which are processed through the somatosensory system, resulting in rich and nuanced flavor experiences that are difficult to model mathematically \cite{gutierrez2021physiology}. Another significant challenge is the limited scope of existing research on gustatory information, especially in comparison to other sensory domains like vision, hearing, and smell. Consequently, capturing the holistic and multisensory nature of taste perception remains a formidable task.

Current approaches to mathematical modeling of the gustatory system can be broadly categorized into two domains: (1) the development of biophysically detailed models that simulate the physiological processes underlying taste perception, and (2) the design of neuromorphic gustatory systems that leverage machine learning techniques and recent advances in microelectronics to replicate or augment gustatory functions. At work \cite{tateno2007network} presents a computational model inspired by the physiological properties of mammalian taste buds. The model consists of a network of taste-sensing cells (modeled as leaky integrate-and-fire neurons) and output cells (bursting neurons). The study explores how noisy inputs from taste-sensing cells can induce synchronization in the bursting cells, with the degree of synchronization serving as a potential indicator of taste stimuli. The results suggest that the network model can detect taste stimuli even without detailed knowledge of the sensing mechanisms, highlighting the role of noise-induced synchronization in taste information processing. The work is supported by physiological findings, including the expression of connexins and ATP receptors in taste bud cells, which facilitate intercellular communication. The research aims to contribute to the development of bio-inspired taste-sensing devices. The study \cite{soltic2008evolving} employs leaky integrate-and-fire (LIF) neurons, a biologically plausible model where postsynaptic potentials summate until a threshold is reached, triggering a spike followed by reset. Inputs are encoded via rank-order population coding (RO-POP C), combining Gaussian receptive fields (for feature resolution) and temporal rank-order spiking (favoring early spikes for salient features), mimicking biological gustatory processing. The evolving spiking neural network (ESNN) learns incrementally: each training pattern either creates a new output neuron or merges with a similar one (based on synaptic weight similarity), enabling one-pass, lifelong learning without catastrophic forgetting—akin to biological neural plasticity. Synaptic weights adapt via a spike-timing-dependent rule, strengthening connections for early spikes, analogous to Hebbian-like reinforcement. While simplified (e.g., no inhibitory synapses, fixed threshold ratio), the model captures key bio-inspired principles: sparse coding, temporal dynamics, and structural adaptability, though it omits detailed biophysical properties (e.g., ion channels). The approach balances computational efficiency with biosimilarity, leveraging the speed of LIF neurons and population coding to approximate the brain’s ability to discriminate tastes via distributed, dynamic spiking patterns.

At work \cite{kim2006simple} presents a portable electronic tongue using a multi-array chemical sensor (MACS) with solid-state ion-selective electrodes (ISEs) and statistical pattern recognition instead of biomimetic approaches. The system employs Principal Component Analysis (PCA) for visual clustering of taste data and Fuzzy C-Means (FCM) for quantitative classification, achieving high accuracy (up to 99\%) in distinguishing beers and teas by origin. Despite its non-biomimetic design, the combination of simple ISEs and machine learning enables robust taste discrimination, nevertheless not giving opportunity of building relations with the biological case. This study presents a deep learning approach to classify molecules into bitter, sweet, and umami taste categories. \cite{dutta2023classification}. Using a curated dataset of 3,706 molecules, researchers developed two models: a descriptor-based a Deep Neural Network (DNN) and a structure-based a Graph Neural Network (GNN). Both achieved ~80-86\% accuracy, with GNNs showing particular promise by learning directly from molecular graphs without requiring handcrafted features. The models successfully screened food compounds, predicting thousands of potential tastants. This work demonstrates AI's potential for tastant discovery in food and pharmaceutical applications, with future potential to expand to all five basic tastes. The approach offers a powerful tool for high-throughput screening of flavor molecules.

Our work advances the field of gustatory modeling by integrating biomorphic design, detailed synaptic dynamics, and multiscale learning mechanisms. First, we implement biomorphic taste cells that incorporate modality-specific receptor dynamics—such as T1R/T2R for sweet and bitter tastes, and ENaC channels for salty stimuli—alongside Goldman–Hodgkin–Katz (GHK)-based ion current calculations to simulate realistic transduction mechanisms. Second, we develop detailed synapse models by incorporating glutamate release kinetics shaped by alpha-function profiles, AMPA receptor trafficking modulated by phosphorylation states, and spike-timing-dependent plasticity (STDP) rules that capture adaptive synaptic behavior. Finally, we introduce a multiscale learning framework that optimizes both temporal coding, using spike-time synchronization metrics, and combinatorial coding, by analyzing population-level firing patterns, enabling robust and biologically plausible taste representation.

\section*{Materials and Methods}

\subsection*{Modeling of Individual Taste Cells}  
The taste cell was modeled as a computational unit integrating four types of taste detectors, each corresponding to a distinct taste modality:  
\begin{itemize}
	\item \textbf{Sweet} detection via \textit{T1R2+T1R3} receptors (metabotropic),  
	\item \textbf{Bitter} detection via \textit{T2R} receptors (metabotropic),  
	\item \textbf{Salty} detection via potential-independent \textit{Na\textsuperscript{+} channels},  
	\item \textbf{Sour} detection via potential-independent \textit{proton channels}.  
\end{itemize}

\subsubsection*{Ionic Current Calculations}  
\begin{enumerate}
	\item \textbf{Electrodiffusion currents (Salty/Sour)}  
	The current through Na\textsuperscript{+} and H\textsuperscript{+} channels was computed using the \textbf{Goldman-Hodgkin-Katz (GHK)} equation for the reversal potential ($E_x$) of each ion:  
	\[
	E_x = \frac{RT}{zF} \ln \left( \frac{[X]_{\text{out}}}{[X]_{\text{in}}} \right),
	\]  
	where $[X]_{\text{out}}$ and $[X]_{\text{in}}$ are extracellular and intracellular ion concentrations, respectively, $R$ is the gas constant, $T$ is temperature, $z$ is ion valence, and $F$ is Faraday’s constant.  
	
	The resultant current ($I_x$) was calculated as:  
	\[
	I_x = g_x (V_m - E_x),
	\]  
	where $g_x$ is the channel conductance and $V_m$ is the membrane potential.  
	
	\item \textbf{Metabotropic currents (Sweet/Bitter)}  
	The activation kinetics of metabotropic receptors (T1R2+T1R3, T2R) were modeled using \textbf{Michaelis-Menten-type equations}, where the fraction of open channels ($O$) depended on tastant concentration ($[L]$) and the half-maximal activation constant ($K_a$), derived from experimentally reported minimal detectable concentrations:  
	\[
	O(t) = \frac{[L]^n}{[L]^n + K_a^n}.
	\]  
	The resulting current ($I_{\text{meta}}$) was computed assuming the reversal potential matched that of voltage-gated Na\textsuperscript{+} channels ($E_{\text{Na}}$):  
	\[
	I_{\text{meta}} = g_{\text{meta}} O(t) (V_m - E_{\text{Na}}).
	\]  
\end{enumerate}

\subsubsection*{Integration into Electrophysiological Dynamics}  
The total membrane current ($I_{\text{tot}}$) was the sum of all ionic and metabotropic contributions:  
\[
I_{\text{tot}} = I_{\text{Na}} + I_{\text{H}} + I_{\text{sweet}} + I_{\text{bitter}}.
\]  
The cellular response was simulated using a hybrid \textbf{Hodgkin-Huxley (HH)} and \textbf{Izhikevich model}, with membrane potential dynamics governed by:  
\[
C_m \frac{dV_m}{dt} = I_{\text{tot}} + I_{\text{pd}},
\]  
where $C_m$ is membrane capacitance and $I_{\text{pd}}$ represents the current through potential-dependent channels (as in the classical model). Spiking behavior was further refined using Izhikevich’s reduced model for action potential generation.  

\subsubsection*{Numerical Integration}  
The differential equations were solved using:  
\begin{itemize}
	\item \textbf{Fourth-order Runge-Kutta (RK4)} for HH dynamics,  
	\item \textbf{Forward Euler} for Izhikevich spiking kinetics.  
\end{itemize}

\subsection*{Post-Synaptic Neuron Model}

\subsubsection*{Spike Train Generation}
The membrane potential response \( V(t) \) of a taste cell was converted into a discrete spike train. Time was partitioned into \( 5\,\text{ms} \) bins, and a binary vector \( \mathbf{S} \in \mathbb{F}_2 \) was constructed such that:
\[
S_k = 
\begin{cases} 
	1 & \text{if } \exists t \in [k\Delta t, (k+1)\Delta t) \text{ where } V(t) \geq V_{\text{threshold}}, \\
	0 & \text{otherwise},
\end{cases}
\]
where \( \Delta t = 5\,\text{ms} \) and \( V_{\text{threshold}} \) is the spike detection threshold.

\subsubsection*{Synaptic Conductance Dynamics}
\begin{enumerate}
	\item \textbf{Alpha Function for Glutamate Release}  
	The time-dependent profile of glutamate release was modeled using a rectified alpha function:
	\[
	\alpha(t) = \begin{cases} 
		\frac{t}{\tau} \exp\left(1 - \frac{t}{\tau}\right) & \text{if } t \geq 0, \\
		0 & \text{otherwise},
	\end{cases}
	\]
	where \( \tau \) is the time constant of glutamate emission.
	
	\item \textbf{Glutamate Concentration in the Synaptic Cleft}  
	The total glutamate concentration \( G(t) \) at time \( t \) was computed as:
	\[
	G(t) = A \sum_{j} \alpha(t - t_j),
	\]
	where \( A \) is the amplitude of glutamate emission per spike, and \( \{t_j\} \) are the presynaptic spike times.
	
	\item \textbf{Synaptic Conductance \( g_{\text{syn}} \)}  
	The total synaptic conductance incorporated phosphorylation-dependent AMPA receptor modulation:
	\begin{equation}
		g_{\text{syn}}(t) = \sum_{i=1}^{N} \left(
		\frac{g_{\text{AMPA}} \cdot N_{\text{total},i} \cdot \text{Phos}_i \cdot G_i(t)}{G_i(t) + K_{\text{Phos}}}
		+ \frac{g_{\text{AMPA}} \cdot N_{\text{total},i} \cdot (1 - \text{Phos}_i) \cdot G_i(t)}{G_i(t) + K_{\text{unPhos}}}
		\right),
	\end{equation}
	where:
	\begin{itemize}
		\item \( g_{\text{AMPA}} \): Maximal conductance of an AMPA receptor,
		\item \( N_{\text{total},i} \): Total number of AMPA receptors at synapse \( i \),
		\item \( \text{Phos}_i \): Phosphorylation rate (\( \in [0,1] \)), i.e the portion of phosphorylated receptors at synapse \( i \),
		\item \( K_{\text{Phos}} \), \( K_{\text{unPhos}} \): Michaelis constants for phosphorylated/unphosphorylated states of the receptor,
		\item \( G_i(t) \): Glutamate concentration at synapse \( i \).
	\end{itemize}
\end{enumerate}

\subsubsection*{Synaptic Current Integration}
The synaptic current \( I_{\text{syn}} \) was computed as:
\[
I_{\text{syn}}(t) = g_{\text{syn}}(t) (V(t) - E_{\text{AMPA}}),
\]
with \( E_{\text{AMPA}} \) being the AMPA receptor reversal potential. This was added to the membrane current in the Hodgkin-Huxley/Izhikevich framework:
\[
C_m \frac{dV}{dt} = -\left(I_{\text{ion}} + I_{\text{syn}}\right) + I_{\text{ext}}.
\]

\subsection*{Global Network Architecture}

\subsubsection*{Network Structure}
Using the described spike encoding and synaptic transmission mechanisms, we constructed a feedforward spiking neural network with the following layers:
\begin{itemize}
	\item \textbf{Input Layer}: Composed of taste cells with varying arrangements of taste detectors (sweet, bitter, salty, sour). Each cell's output was encoded as a binary spike train $S_k \in \mathbb{F}_2$ using the $5\,\text{ms}$ binning procedure described in Section~2.2.
	
	\item \textbf{Hidden Layers}: Consisting of post-synaptic neurons driven by glutamate-mediated synapses. The synaptic dynamics followed the phosphorylation-dependent conductance model:
	\[
	g_{\text{syn},i}(t) = \frac{g_{\text{AMPA}} N_{\text{total},i} G_i(t)}{G_i(t) + K_{\text{eff}}}
	\]
	where $K_{\text{eff}} = K_{\text{Phos}}$ (if phosphorylated) or $K_{\text{unPhos}}$ (otherwise), implementing activity-dependent plasticity.
\end{itemize}

\subsubsection*{Weight Representation}
The network's synaptic weights were determined by two physiological parameters:
\begin{enumerate}
	\item \textit{Total AMPA receptors} ($N_{\text{total},i}$): Governed baseline connection strength.
	\item \textit{Phosphorylation rate} ($\text{Phos}_i$): Modulated efficacy via long-term potentiation (LTP), with updates following:
	\[
	\Delta \text{Phos}_i =  \eta \cdot \sum_{j} S_j \cdot (1 - \text{Phos}_i) / N,
	\]
	
	where:
	\begin{itemize}
		\item $\eta$: Learning rate (fixed at 0.05)
		\item $S_j$: Synchronization index for spike train pair $j$, defined as:
		\[
		S_j = 0.5 - \frac{d_j + \Delta t}{D}
		\]
		\begin{itemize}
			\item $d_j$: Minimal interspike interval between pre- and post-synaptic neurons in trial $j$
			\item $\Delta t$: Temporal resolution bin size (5 ms)
			\item $D$: Maximum observed interspike interval across all trials
		\end{itemize}
		\item N: The number of spike train pairs.
	\end{itemize}
	
\end{enumerate}

\subsubsection*{Output Encoding}
The network produced a $k \times m$-dimensional output vector over response period $T = 300\,\text{ms}$:
\[
\mathbf{Y} \in \mathbb{F}_2^{k \times m}, \quad k = \left\lfloor \frac{T}{\Delta t} \right\rfloor
\]
where:
\begin{itemize}
	\item $k = 60$ bins ($\Delta t = 5\,\text{ms}$),
	\item $m$: Number of output neurons,
	\item Each entry $Y_{i,j}$ indicated a spike in neuron $j$ during bin $i$.
\end{itemize}

\subsubsection*{Information Flow}
\begin{enumerate}
	\item Taste inputs $\rightarrow$ Spike trains via receptor dynamics (Section~2.1),
	\item Spike trains $\rightarrow$ Synaptic currents (Section~2.2),
	\item Currents $\rightarrow$ Output spikes through iterative membrane potential updates.
\end{enumerate}

\subsection*{Learning}

\subsubsection*{Genetic Optimization}
The network's synaptic weights were optimized through an evolutionary approach:
\begin{itemize}
	\item \textbf{AMPA} Receptor Counts $N_{\text{total},i}$: Optimized using a genetic algorithm with:
	\begin{itemize}
		\item Population size: 100 individuals
		\item Selection: Rank-based, keeping top 50\% performers
		\item Mutation: Gaussian noise ($\sigma = 0.1N_{\text{total},i}$)
		\item Crossover: Uniform mixing of parent parameters
	\end{itemize}
	
	\item \textbf{Phosphorylation Rates ($\text{Phos}_i$)}: Adapted online according to synaptic activity:
	\[
	\text{Phos}_i^{(n+1)} = \text{Phos}_i^{(n)} + \Delta \text{Phos}_i^{\text{(top 50\%)}}
	\]
	where $\Delta \text{Phos}_i^{\text{(top 50\%)}}$ is the mean $\Delta \text{Phos}_i$ in the best-performing individuals.
\end{itemize}

\subsection*{Parameter Values}
All model constants were derived from experimental data or calibrated through simulation:

\begin{table}[h]
	\centering
	\caption{Key Model Parameters}
	\begin{tabular}{lll}
		\hline
		\textbf{Parameter} & \textbf{Value} & \textbf{Description} \\
		\hline
		$\tau$ & 2.5\,\text{ms} & Glutamate emission time constant \\
		$A$ & 1.2\,\text{mM} & Glutamate release amplitude \\
		$g_{\text{AMPA}}$ & 10\,\text{pS} & Single AMPA receptor conductance \\
		$K_{\text{Phos}}$ & 0.3\,\text{mM} & Phosphorylated AMPA Michaelis constant \\
		$K_{\text{unPhos}}$ & 1.5\,\text{mM} & Unphosphorylated AMPA Michaelis constant \\
		$E_{\text{AMPAR}}$ & 0\,\text{mV} & AMPA reversal potential \\
		$\eta$ & 0.05 & Phosphorylation learning rate \\
		$V_{\text{threshold}}$ & -50\,\text{mV} & Spike detection threshold \\
		\hline
	\end{tabular}
\end{table}

\begin{itemize}
	\item Temperature: $T = 310\,\text{K}$ (physiological)
	\item Bin width: $\Delta t = 5\,\text{ms}$ (temporal resolution)
	\item Simulation time: $T = 300\,\text{ms}$ per trial
\end{itemize}

\section*{Results}

Firstly, we need to note that training was done only for the Izhikevich version of the model due to its computational efficiency compared to the full Hodgkin-Huxley dynamics. Therefore, all subsequent results and discussions pertain to the Izhikevich-based network.

\subsection*{Experimental Procedure and Network Evaluation}
To evaluate the network's ability to encode and discriminate taste stimuli, we conducted a series of computational experiments. Each experiment involved presenting the model with a specific taste stimulus, represented as a vector of four concentrations corresponding to the primary taste modalities: salty ($[Na^+]$), sour ($[H^+]$), sweet, and bitter. 

As described in the Methods section, these input concentrations modulate the ionic currents in the input layer neurons, causing changes in their membrane potentials. When a neuron's potential crosses the firing threshold ($V_{\text{threshold}}$), it generates a spike. The collective spiking activity of all neurons across all layers was simulated for a period of 300 ms for each input stimulus. The resulting output of the model for a given stimulus is a set of spike raster plots, which show the precise timing of spikes for each neuron in the network over the simulation period. The training process, driven by a genetic algorithm, aimed to optimize the network's synaptic weights to produce distinct and informative output spike patterns for different tastes, based on the specific loss functions defined for each task.

\subsection*{Untrained Network Responses}
Before training, we analyzed the network's baseline activity in response to taste stimuli. Figures 2 and 3 show the spike raster plots for all layers of the untrained network when presented with a mid-concentration salty input and a high-concentration mixed sweet-sour input, respectively. The activity across all layers appears dense, noisy, and lacks clear, discernible patterns. This unstructured response serves as a control, demonstrating that without optimization, the network is incapable of forming a specific representation of the taste stimulus.

\subsection*{Learning Dynamics}
The network was trained on two distinct tasks: encoding the perceived pleasantness (hedonic value) of tastes and discriminating between pure and mixed taste stimuli. The learning progress was tracked by monitoring the fitness function, which corresponds to the negative of the loss function, maximized by the genetic algorithm.

Figures 4 and 5 illustrate the evolution of the best fitness score across generations for Task 1 and Task 2, respectively. In both cases, the fitness value steadily increases and converges, indicating that the genetic algorithm successfully optimized the network's synaptic parameters (AMPA receptor counts and phosphorylation rates) to minimize the error between the target taste representation and the network's output spike patterns. This confirms that the model is capable of learning the desired input-output mappings.

\subsection*{Trained Network Responses and Taste Recognition}
After the training process was completed, the network was re-evaluated using the same input stimuli as in the pre-training phase. The resulting spike patterns for the trained network are shown in Figure 6 (mid-concentration salty) and Figure 7 (high-concentration mixed sweet-sour). 

In sharp contrast to the chaotic activity of the untrained network, the responses of the trained network are significantly more sparse, structured, and reproducible. Specific groups of neurons fire in coordinated, temporally precise patterns that are unique to the input stimulus. This transformation from noisy to structured activity demonstrates that the network has learned to form distinct and efficient neural codes for different tastes. The emergence of these specific output patterns is the basis for taste recognition, as they represent a unique "neural fingerprint" for each stimulus, which can be reliably decoded by downstream neural structures to identify the taste.
\subsubsection*{Task 1: Encoding Taste Pleasantness}
The loss function quantified the preservation of hedonic distance between stimuli:
\[
\mathcal{L}_1 = \sum_{i=1}^{n}-\frac{|D_{\text{in}}(A_i,B_i) - D_{\text{res}}(A_i,B_i)|}{n}
\]
where:
\begin{itemize}
    \item $D_{\text{taste}}(A,B)$: Distance in taste pleasantness space for stimuli $A=(A_{\text{Na}}, A_{\text{H}}, A_{\text{S}}, A_{\text{Q}})$ and $B=(B_{\text{Na}}, B_{\text{H}}, B_{\text{S}}, B_{\text{Q}})$:
    \[
    D_{\text{taste}}(A,B) = \left| \prod_{i\in\{\text{Na,H,S,Q}\}} \exp\left(-\frac{(A_i - \theta_i)^2}{w_i^2}\right) - \prod_{i\in\{\text{Na,H,S,Q}\}} \exp\left(-\frac{(B_i - \theta_i)^2}{w_i^2}\right) \right|
    \]
    where:
    \begin{itemize}
        \item $\theta_{\text{Na}} = 100.0$, $w_{\text{Na}} = 50.0$ (optimal NaCl concentration and width)
        \item $\theta_{\text{H}} = 0.1$, $w_{\text{H}} = 2.0$ (optimal acidity and width)
        \item $\theta_{\text{S}} = 15000.0$, $w_{\text{S}} = 10000.0$ (optimal sweetness and width)
        \item $\theta_{\text{Q}} = 1.0$, $w_{\text{Q}} = 5.0$ (optimal bitterness and width)
    \end{itemize}
    
    \item $D_{\text{res}}$: Information distance between output spike trains:
    \[
    D_{\text{res}}(A,B) = H(X,Y) - MI(X;Y)
    \]
    with $H$ as joint entropy and $MI$ as mutual information.
\end{itemize}

\subsubsection*{Task 2: Taste Discrimination}
For pure vs. mixed taste differentiation, the input metric was:
\[
D_{\text{in}}(A,B) = 
\begin{cases}
    \|A\|_1 + \|B\|_1 & \text{if both pure tastes} \\
    \frac{1}{2}\|A - B\|_2 & \text{otherwise}
\end{cases}
\]
with output distance $D_{\text{res}}$ computed as in Task 1. The loss:
\[
\mathcal{L}_2 = \sum_{i=1}^{n}-\frac{|D_{\text{in}}(A_i,B_i) - D_{\text{res}}(A_i,B_i)|}{n}
\]

\section*{Discussion}

The hybrid model of taste perception presented in this work successfully integrates biological fidelity at the level of ion channels and synaptic transmission with the computational efficiency required for medium-scale network simulations. Our model demonstrates how the optimization of synaptic parameters via a genetic algorithm enables the network to form sparse and structured spike patterns that serve as unique ``neural fingerprints'' for different taste stimuli. This serves as a proof of principle that biomorphic computing can efficiently encode complex sensory information.

\textbf{Limitations of the Model}

Despite the promising results, our model has several limitations that are important to acknowledge.

\begin{enumerate}
    \item \textbf{Biophysical Simplification:} Although we utilized GHK and Michaelis-Menten equations to describe currents, the model omits many intricacies of intracellular signal transduction. For instance, for bitter, sweet, and umami tastes, the cascades of secondary messengers (IP\textsubscript{3}, DAG), the dynamics of Ca\textsuperscript{2+} release from intracellular stores, and the subtle modulation of voltage-gated channels are not fully modeled. This may limit the accuracy of predicting responses to complex or low-concentration stimuli.

    \item \textbf{Network Architecture:} The feedforward architecture used is a significant simplification compared to the real gustatory system, which features abundant feedback loops both within taste buds and at the level of the brainstem and thalamus. These feedback connections play a crucial role in adaptation, habituation, and the modulation of taste perception by other modalities (e.g., olfaction).

    \item \textbf{Influence of Other Sensory Modalities:} As noted in the introduction, taste is closely integrated with olfaction, somatosensory (texture, temperature), and even visual sensations. Our current model is purely gustatory and does not account for this multisensory integration, which is fundamental for forming the holistic perception of ``flavor''.

    \item \textbf{Learning and Plasticity:} Although we implemented an STDP-like rule for AMPA receptor phosphorylation, the learning mechanism was global and driven by an external genetic algorithm. In a real biological system, plasticity is local and distributed. Furthermore, the model lacks inhibitory interneurons, which are critical for forming contrastive and selective neural representations.
\end{enumerate}

\textbf{Future Directions and Model Development}

Overcoming these limitations opens several fruitful avenues for future research:

\begin{enumerate}
    \item \textbf{Multisensory Integration:} The most evident development is the integration of an olfactory model. Following the same paradigm, a biomorphic model of the olfactory bulb could be created, where information is encoded by spatiotemporal spike patterns in glomeruli and mitral cells, and its output could be connected to the gustatory network. This would allow for the study of how combined taste-odor stimuli are encoded and discriminated at higher processing levels.

    \item \textbf{Incorporating Feedback and Inhibition:} Extending the network architecture by including recurrent connections and populations of inhibitory interneurons (e.g., based on the Izhikevich model) would enable more complex and robust dynamic regimes, such as synchronization and competitive interaction (``winner-take-all''), bringing the model closer to its biological prototype.

    \item \textbf{Deepening Intracellular Signaling:} The model can be made more detailed by incorporating comprehensive models of calcium dynamics and G-protein cascades for metabotropic receptors, using stochastic or deterministic systems of differential equations.

\end{enumerate}

\section*{Conclusion}

The development of computationally efficient yet biologically plausible models of taste perception represents a significant challenge at the intersection of computational neuroscience and neuromorphic engineering. This work addresses this challenge by proposing a novel hybrid modeling framework that bridges the gap between detailed biophysical representation and large-scale network simulation. 

Our approach successfully integrates modality-specific receptor dynamics, Goldman-Hodgkin-Katz ion current calculations, and detailed synaptic mechanisms featuring phosphorylation-dependent AMPA receptor trafficking and spike-timing-dependent plasticity. By combining the biological fidelity of Hodgkin-Huxley formalism with the computational efficiency of Izhikevich spiking neurons, we have created a scalable model capable of simulating gustatory transduction and neural coding processes.

The key results demonstrate that our optimized network can transform unstructured input representations into distinct, sparse, and reproducible spiking patterns that serve as neural fingerprints for different taste qualities. The genetic algorithm-based learning effectively tuned synaptic parameters to encode both taste pleasantness and discriminate between pure and mixed stimuli, showcasing the model's capability for multimodal taste representation.

The importance of this research extends beyond theoretical neuroscience, offering practical implications for the development of neuromorphic gustatory systems. The model's architecture provides a foundation for building energy-efficient electronic tongues capable of real-time taste analysis with biological fidelity. Furthermore, the framework establishes a basis for future integration with other sensory modalities, particularly olfaction \cite{stasenko2023model}, toward creating comprehensive flavor perception systems.

While acknowledging current limitations in biophysical detail and network complexity, this work establishes a robust foundation for future developments in computational taste modeling. The proposed approach opens new possibilities for both understanding the neural basis of taste perception and engineering novel bio-inspired computing systems for chemical sensing applications.

\bibliography{sample}

\section*{Acknowledgements}

This work was funded by the $\ldots$.

\section*{Author contributions statement}

V.A.L designed the research and idea. V.A.L simulated the model. V.A.L, S.V.S, and V.B.K, performed data analysis. V.A.L, S.V.S, and V.B.K interpreted the results.  V.A.L, S.V.S, and V.B.K formulated the model. All authors participated in writing and editing the manuscript.

\section*{Competing interests}
The authors declare no competing interests.

\section*{Code availability}
Code used to produce the results presented herein is available from the corresponding author on reasonable request.

\section*{Data availability}
The datasets used and/or analysed during the current study available from the corresponding author on reasonable request.

\section*{Additional information}

Correspondence and requests for materials should be addressed to S.V.S. 

\begin{figure}[h!]
	
	\centering
	\textit{(a)}\includegraphics[width=0.45\textwidth]{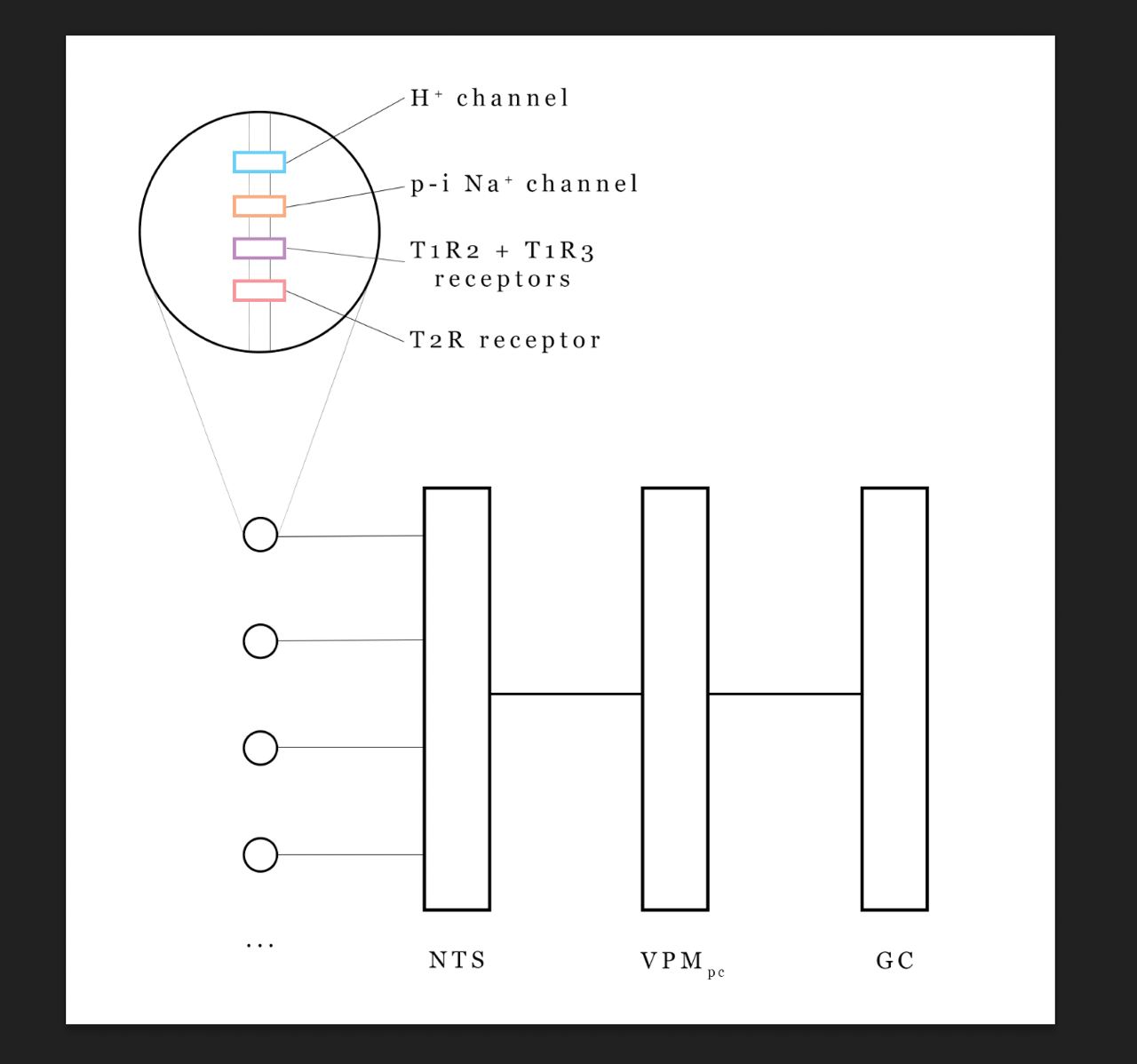}
	\textit{(b)}\includegraphics[width=0.45\textwidth]{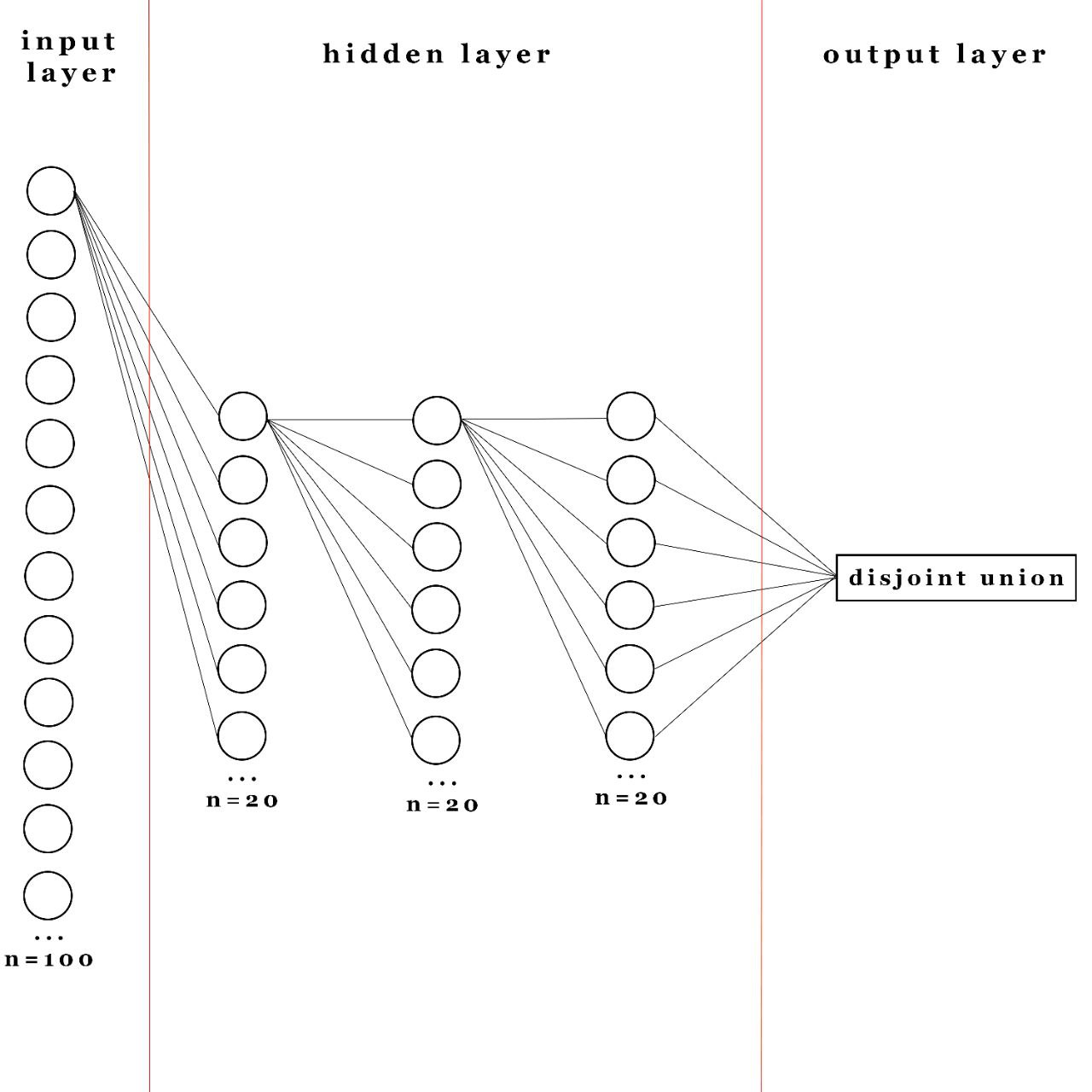}
	\caption{ a) The scheme of mamalian taste analyzer. b) The biological taste analyzer and our model structure comparison.}
	\label{fig:both_images}\vspace{5pt}
\end{figure}

\begin{figure}[h!]
    \centering
\includegraphics[width=1\linewidth]{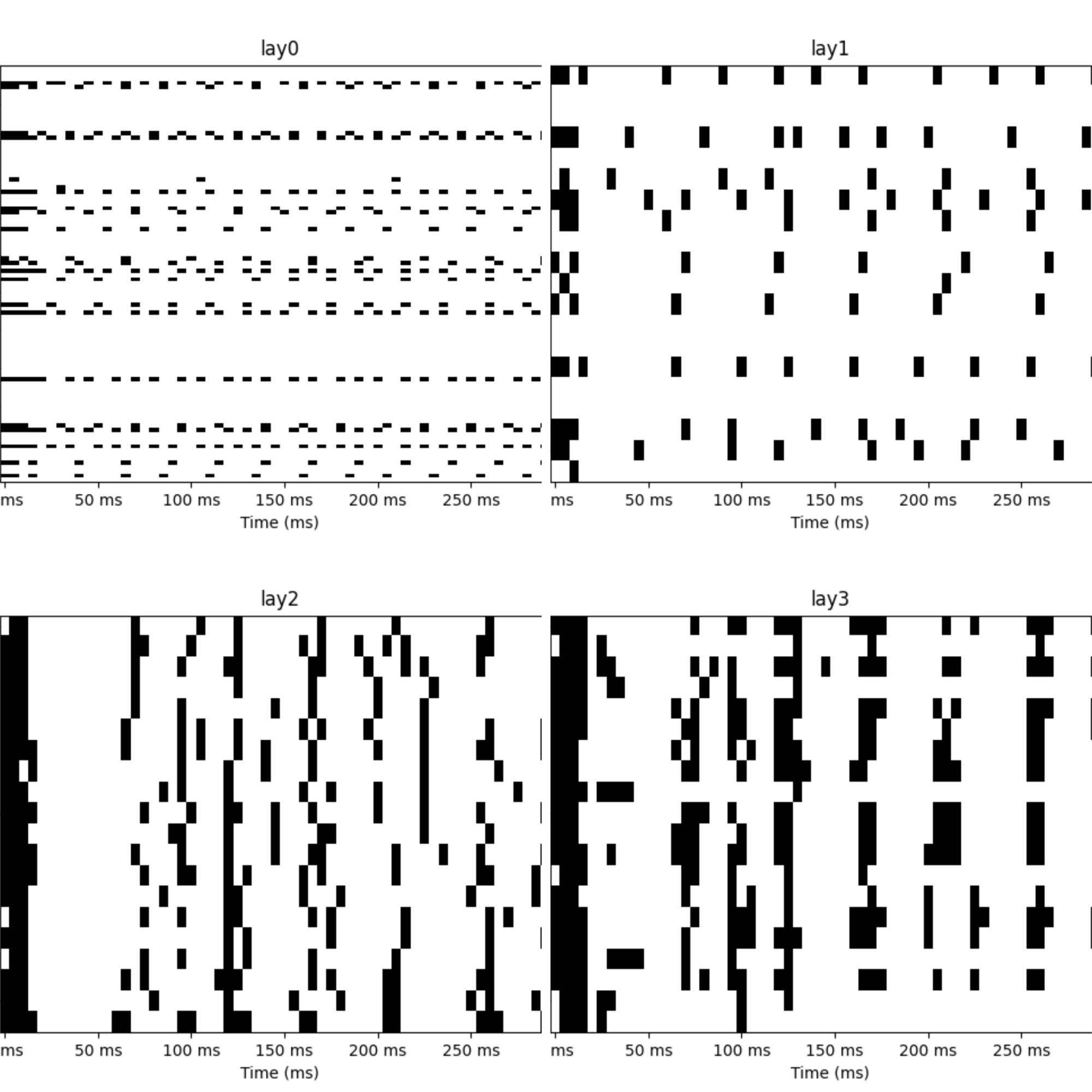}
\caption{Raw output spike patterns of all the untrained network's layers to a mid-concentraited salty input. Each row shows the spike raster of a taste cell population over 300 ms.}
\label{fig:untrained}
\end{figure}

\begin{figure}[h!]
    \centering
\includegraphics[width=1\linewidth]{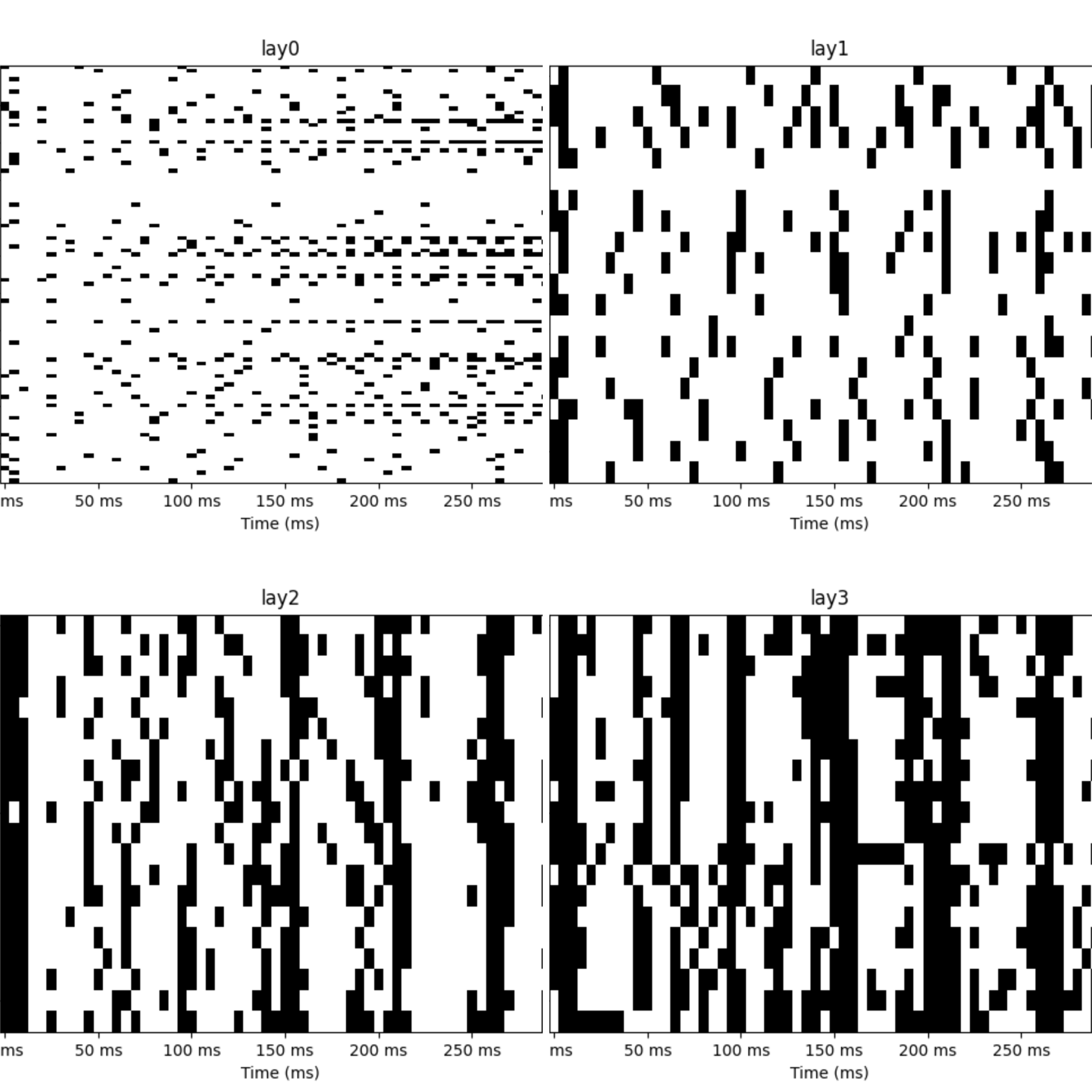}
\caption{Raw output spike patterns of all the untrained network's layers to a high-concentraited mixed sweet-sour input. Each row shows the spike raster of a taste cell population over 300 ms.}
\label{fig:untrained2}
\end{figure}

\begin{figure}[h!]
    \centering
\includegraphics[width=0.9\linewidth]{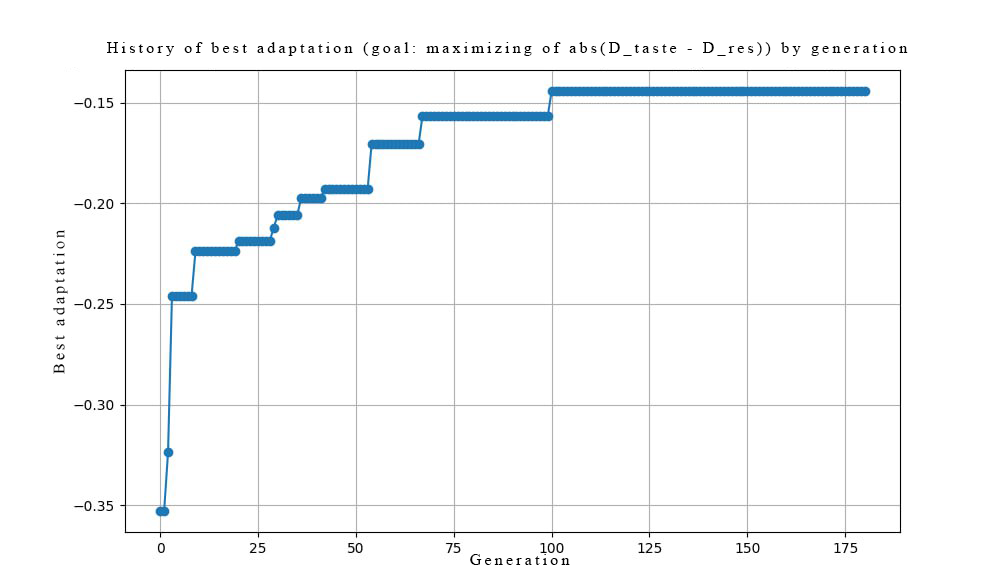}
\caption{Evolution of fitness (maximizing $|D_{\text{taste}} - D_{\text{res}}|$) across generations. Normalized to $[-1,0]$ range.}
\label{fig:pleasure}
\end{figure}

\begin{figure}[h!]
    \centering
\includegraphics[width=0.9\linewidth]{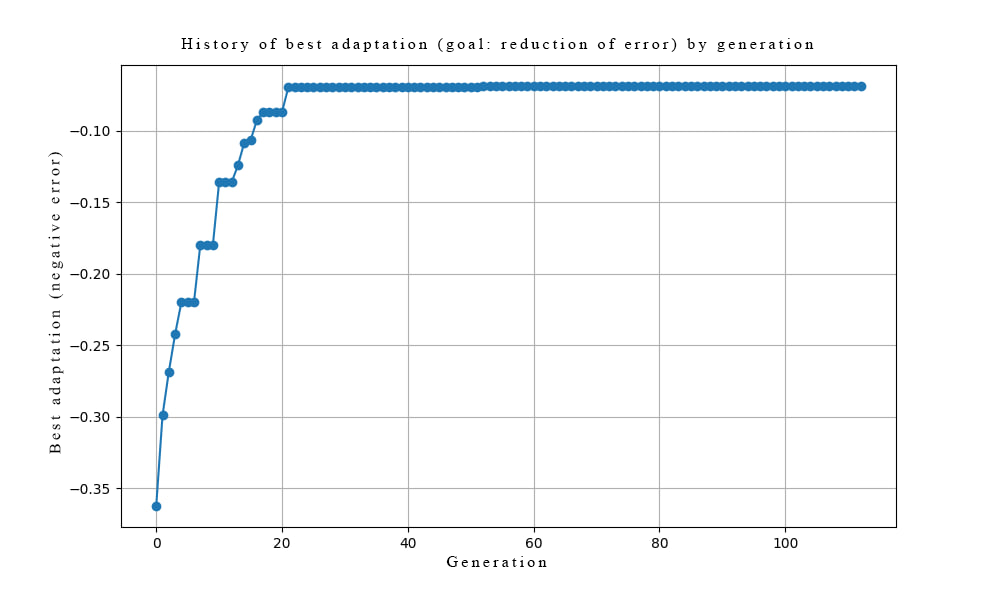}
\caption{Evolution of fitness (maximizing $|D_{\text{taste}} - D_{\text{res}}|$) across generations. Normalized to $[-1,0]$ range.}
\label{fig:discrimination}
\end{figure}

\begin{figure}[h!]
    \centering
\includegraphics[width=1\linewidth]{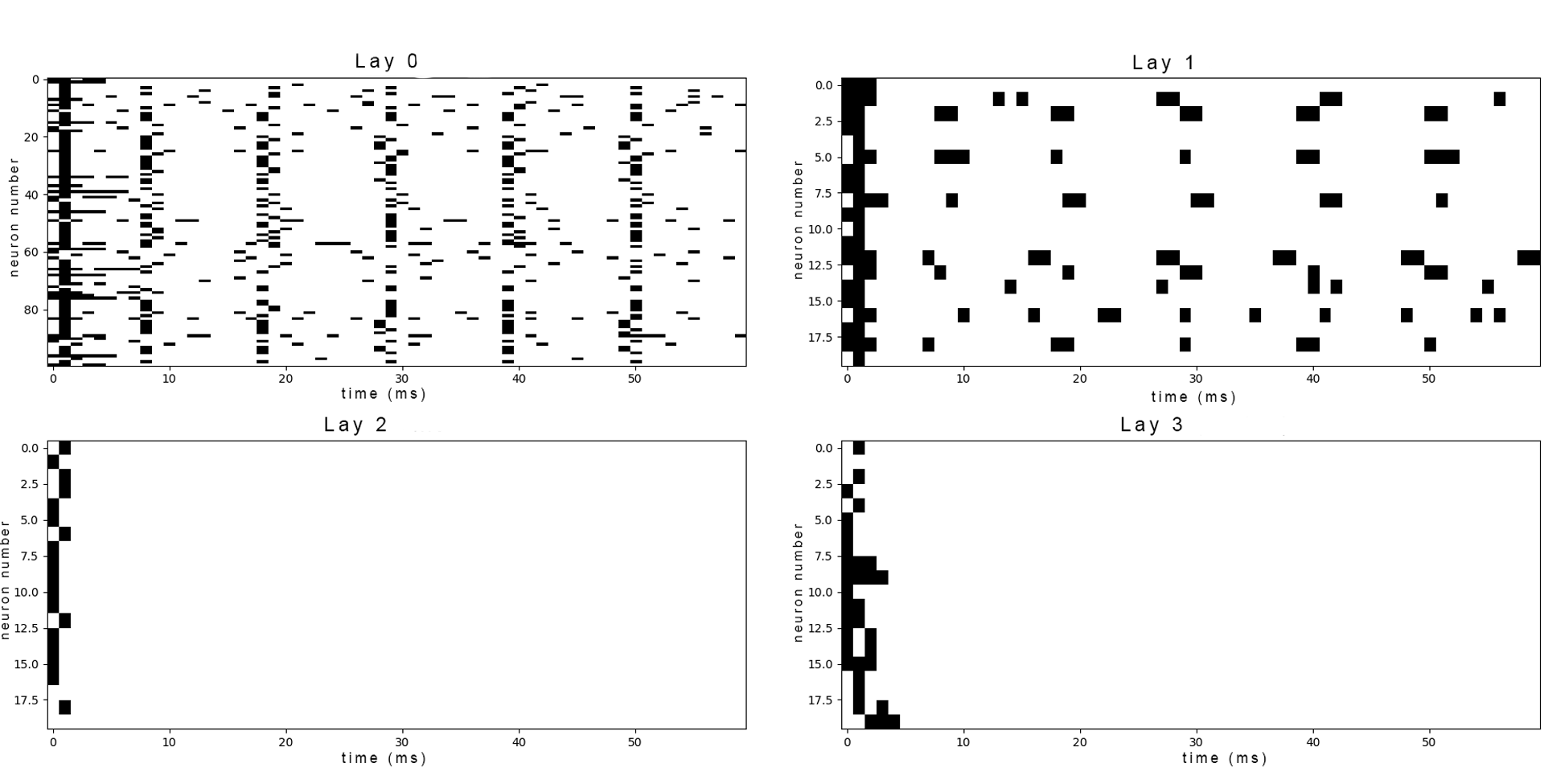}
\caption{Raw output spike patterns of all the trained network's layers to a mid-concentraited salty input. Each row shows the spike raster of a taste cell population over 300 ms.}
\label{fig:untrained3}
\end{figure}

\begin{figure}[h!]
    \centering
\includegraphics[width=1\linewidth]{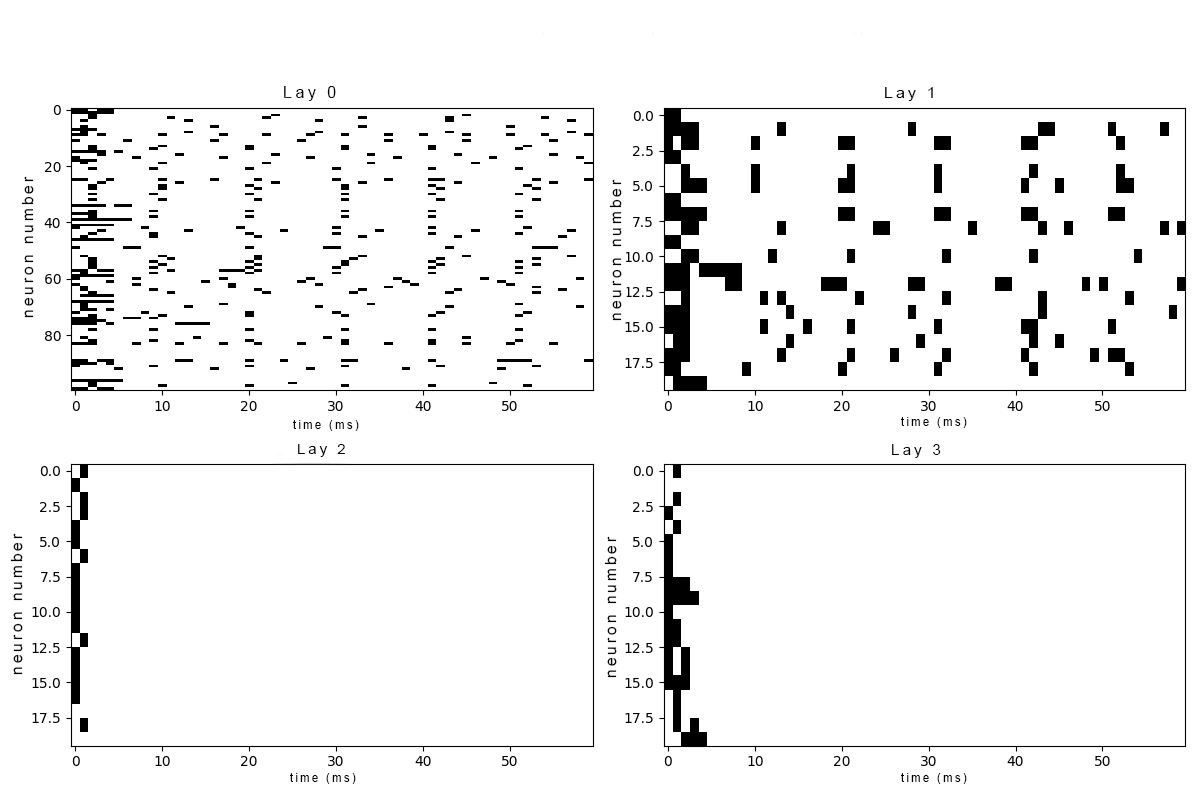}
\caption{Raw output spike patterns of all the trained network's layers to a high-concentraited mixed sweet-sour input. Each row shows the spike raster of a taste cell population over 300 ms.}
\label{fig:untrained4}
\end{figure}

\end{document}